\documentclass[8.5pt,twoside,twocolumn]{article}
\usepackage[super,sort&compress,comma]{natbib} 
\usepackage[version=3]{mhchem}
\usepackage{balance}
\usepackage{times,mathptm}
\usepackage{sectsty}
\usepackage{graphicx} 
\usepackage{lastpage}
\usepackage[format=plain,justification=raggedright,singlelinecheck=false,font=small,labelfont=bf,labelsep=space]{caption} 
\usepackage{fancyhdr}
\begin{document}

\thispagestyle{plain}
\fancypagestyle{plain}{

%\fancyhead[L]{\includegraphics[height=8pt]{headers/LH}}
%\fancyhead[C]{\hspace{-1cm}\includegraphics[height=20pt]{headers/CH}}
%\fancyhead[R]{\hspace{10cm}\vspace{-0.25cm}\includegraphics[height=10pt]{headers/RH}}

\renewcommand{\headrulewidth}{1pt}}
\renewcommand{\thefootnote}{\fnsymbol{footnote}}
\renewcommand\footnoterule{\vspace*{1pt}% 
\hrule width 3.4in height 0.4pt \vspace*{5pt}} 
\setcounter{secnumdepth}{5}

\makeatletter 
\renewcommand\@biblabel[1]{#1}            
\renewcommand\@makefntext[1]% 
{\noindent\makebox[0pt][r]{\@thefnmark\,}#1}
\makeatother 
\renewcommand{\figurename}{\small{Fig.}~}
\sectionfont{\large}
\subsectionfont{\normalsize} 

\fancyfoot{}

%\fancyfoot[LO,RE]{\vspace{-7pt}\includegraphics[height=9pt]{headers/LF}}
%\fancyfoot[CO]{\vspace{-7.2pt}\hspace{12.2cm}\includegraphics{headers/RF}}
%\fancyfoot[CE]{\vspace{-7.5pt}\hspace{-13.5cm}\includegraphics{headers/RF}}

%\fancyfoot[RO]{\footnotesize{\sffamily{1--\pageref{LastPage} ~\textbar  \hspace{2pt}\thepage}}}
%\fancyfoot[LE]{\footnotesize{\sffamily{\thepage~\textbar\hspace{3.45cm} 1--\pageref{LastPage}}}}
%\fancyfoot[LE]{\footnotesize{\sffamily{\thepage~\textbar\hspace{3.45cm} 1--\pageref{LastPage}}}}

\fancyfoot[R]{\footnotesize{\sffamily{\thepage /4}}}
%\fancyfoot[LE]{\footnotesize{\sffamily{\thepage~\textbar\hspace{3.45cm} 1--\pageref{LastPage}}}}
%\fancyfoot[LE]{\footnotesize{\sffamily{\thepage/4}}}

\fancyhead{}
\renewcommand{\headrulewidth}{1pt} 
\renewcommand{\footrulewidth}{1pt}
\setlength{\arrayrulewidth}{1pt}
\setlength{\columnsep}{6.5mm}
\setlength\bibsep{1pt}

\twocolumn[
  \begin{@twocolumnfalse}
\noindent\LARGE{\textbf{The microfluidic Kelvin water dropper}}
\vspace{0.6cm}

\noindent\large{\textbf{\'Alvaro G. Mar\'in,$^{\ast}$\textit{$^{a\ddag}$} Wim van Hoeve,\textit{$^{b}$} Pablo Garc\'ia-S\'anchez,\textit{$^{b}$} Lingling Shui, \textit{$^{c}$} Yanbo Xie,\textit{$^{c}$} Marco A. Fontelos,\textit{$^{d}$} Jan C. T. Eijkel, \textit{$^{c}$} Albert van den Berg,\textit{$^{c}$} and Detlef Lohse\textit{$^{a}$}}}\vspace{0.5cm}
%Please note that \ast indicates the corresponding author(s) but no footnote text is required. 

\noindent\textit{\small{\textbf{Received 12 Jul 2013, Accepted 09 Sep 2013\newline
First published on the web 10 Sep 2013}}}

\noindent \textbf{\small{DOI:10.1039/C3LC50832C.}}
 \end{@twocolumnfalse} \vspace{0.6cm}

  ]

\noindent\textbf{The so-called ``Kelvin water dropper'' is a simple experiment demonstrating the spontaneous appearance of induced free charge in droplets emitted through a tube. As Lord Kelvin explained, water droplets spontaneously acquire a net charge during detachment from a faucet due to the presence of electrical fields in their surrounding created by any metallic object. In his experiment, two streams of droplets are allowed to drip from separated nozzles into separated buckets, which are at the same time interconnected through the dripping needles. In this paper we build a microfluidic water dropper and demonstrate that the droplets get charged and break-up due to electrohydrodynamic instabilities. A comparison with recent simulations shows the dependence of the acquired charge in the droplets on different parameters of the system. The phenomenon opens a door to cheap and accessible transformation of pneumatic pressure into electrical energy and to an enhanced control in microfluidic and biophysical manipulation of capsules, cells and droplets via self-induced charging of the elements.}
\section*{}
\vspace{-1cm}
%Footnotes
%\footnotetext{\dag~Electronic Supplementary Information (ESI) available: [details of any supplementary information available should be included here]. See DOI: 10.1039/b000000x/}

%Please use \dag to cite the ESI in the main text of the article.
%If you article does not have ESI please remove the the \dag symbol from the title and the above footnotetext.

\footnotetext{\textit{$^{a}$Physics of Fluids, University of Twente, Enschede, The Netherlands.}}
\footnotetext{\textit{$^{b}$Depto. de Electr\'onica y Electromagnetismo. Universidad de Sevilla. Spain. }}
\footnotetext{\textit{$^{c}$BIOS/Lab on a chip group, MESA+, Institute of Nanotechnology, University of Twente. The Netherlands}}
\footnotetext{\textit{$^{d}$ Instituto de Ciencias Matem\'aticas, ICMAT, Madrid, Spain}}

%additional addresses can be cited as above using the lower-case letters, c, d, e... If all authors are from the same address, no letter is required

%\footnotetext{\ddag~Additional footnotes to the title and authors can be included \emph{e.g.}\ `Present address:' or `These authors contributed equally to this work' as above using the symbols: \ddag, \textsection, and \P. Please place the appropriate symbol next to the author's name and include a \texttt{\textbackslash footnotetext} entry in the the correct place in the list.}

\footnotetext{\ddag Present address: Instit\"ut f\"ur Str\"omungsmechanik und Aerodynamik, Bundeswehr University Munich, Germany. E-mail: a.marin@unibw.de}

%The main text of the article should appear here.\cite{Mena2000} Communications do not normally have section headings.

\begin{figure}[t!]
\begin{center}
\includegraphics[width=0.4\textwidth]{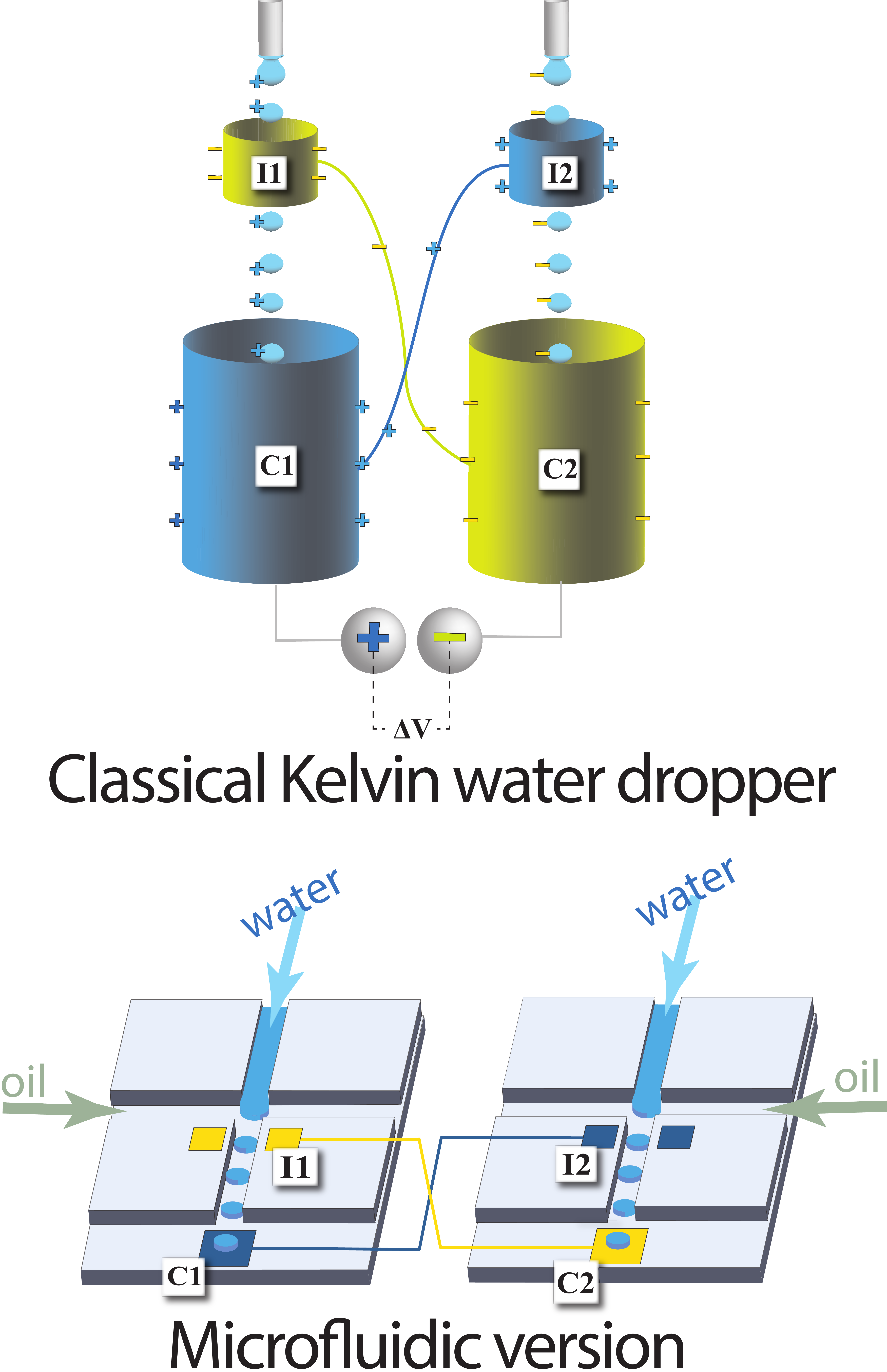}
\end{center}
\caption{Upper sketch: Illustration of a classical Kelvin water dropper as used in popular demonstrations and house-made experiments. Lower sketch: Representation of the microfluidic Kelvin water dropper.\label{fig1}}

\end{figure}

In 1867, Sir William Thomson (later known as Lord Kelvin) devised an apparatus to ``illustrate the voltaic theory'' as he literally stated \cite{Thomson:1867ws}. The apparatus has become very popular as a simple demonstration of electrostatic processes, since it can safely generate high voltages and electrical discharges by just letting water falling from a couple of faucets. The physical mechanism is however complex and still intriguing: two faucets are dripping water into two separate metallic buckets (figure \ref{fig1}) which have to be connected in a particular way. When a drop detaches from a metallic faucet at high enough frequency it will acquire a tiny residual amount of charge. The amount of acquired charge depends on the material of the faucet, on the water electrical conductivity and on the  local electrical field at the detachment point. A metallic ring is now placed under the faucet (I1 in figure \ref{fig1}), so when a droplet passes with a small charge, it will induce an opposite charge in the metallic ring. The droplet continues and ends up in the metallic bucket (C1 in figure \ref{fig1}), with its tiny charge dispersed in it. The idea of Lord Kelvin was to make a self-feeding system with the help of the second faucet: the metallic ring is connected to a second metallic bucket (C2 in figure \ref{fig1}). Therefore, the ring will be slightly charged with the charge induced in the second faucet. Assuming it is negatively charged, the ring will then induce positive charges in the droplets of the first faucet. They will fall down with their positive charges into bucket I1. The bucket is connected to the second metallic ring (I2 in figure \ref{fig1}) in the adjacent system, in such a way that negative charge induction is enhanced in the second faucet's droplets (see figure \ref{fig1}). A voltage difference of order of several kilovolts can be created between the two buckets in a matter of seconds. In popular demonstrations like the classical videos from Melcher, Zahn and Silva \cite{Electricfieldsand:wo} or the more recent by Walter Lewin in MIT \cite{WalterMIT}, electrical sparks can be visible between the two metallic buckets with simple arrangements. 
%Such type of phenomena could be classified as Òelectrical influenceÓÉ similar to the classic Wimhurst machine [find description in Wikipedia], in which two rotating disks develop charge ``spontaneously'' after forcing them to rotate. 
With this experiment, Kelvin intended to explain the spontaneous generation of electricity in atmospheric phenomena as thunderstorms. Besides the unfruitful efforts to apply the invention to obtain AC electrical energy \cite{Zahn:1973wj}, the brilliant device of Lord Kelvin has merely remained since then as an interesting demonstration. 

% \section { Recent progress on micro-electro-fluidics }
 
A way to improve the energy conversion efficiency would be to scale down the apparatus, transforming it into a microfluidic device. Such an approach to increase energy transformation efficiencies has been followed by Xie et al. \cite{Xie2013} in a different system, in which the authors have reported energy conversion from pneumatic into electrical energy with efficiencies up to $48\% $ making use of a charged liquid microjet at high velocity. In other sort of applications as flow cytometry, electrostatic charging is employed for sorting droplets containing cells in special type of FACS (Fluorescent Activated Cell Sorting) \cite{shapiro2003}, and has recently been extended to use dielectrophoretic forces to sort uncharged droplets \cite{Baret:2009wx}.

%\begin{figure}[h!]
%\begin{center}
%\includegraphics[width=.3\textwidth]{figs/fig2.pdf}
%\end{center}
%\caption{Representation of the microfluidic Kelvin water dropper.}
%\label{fig2}
%\end{figure}

In this paper we report the miniaturization of the popular Kelvin water dropper into a microfluidic device, which is able to charge droplets of several picoliters ``spontaneously'' at typical rates of $10^3$ droplets per second. An illustration of the device is depicted in the lower figure \ref{fig1}. It basically consists in two microfluidic drop generators build in PDMS and closed with glass slides, where the electrodes are precisely deposited. The droplets then encounter an induction electrode (analogous to the rings in the classical water dropper), and further downstream a second larger electrode is used to receive the droplet charge (analogous to the metallic bucket in the classical configuration). Inductors and receivers are interconnected to produce the charge buildup. Due to the growth of the voltages, it is crucial to maintain both electrical circuits well isolated and separated to avoid discharges. Also note that the only electrodes in direct contact with the water droplets are the receiving electrodes (C1 and C2 in figure \ref{fig1}) and not the inductors.

\begin{figure}
\begin{center}
\includegraphics[width=.5\textwidth]{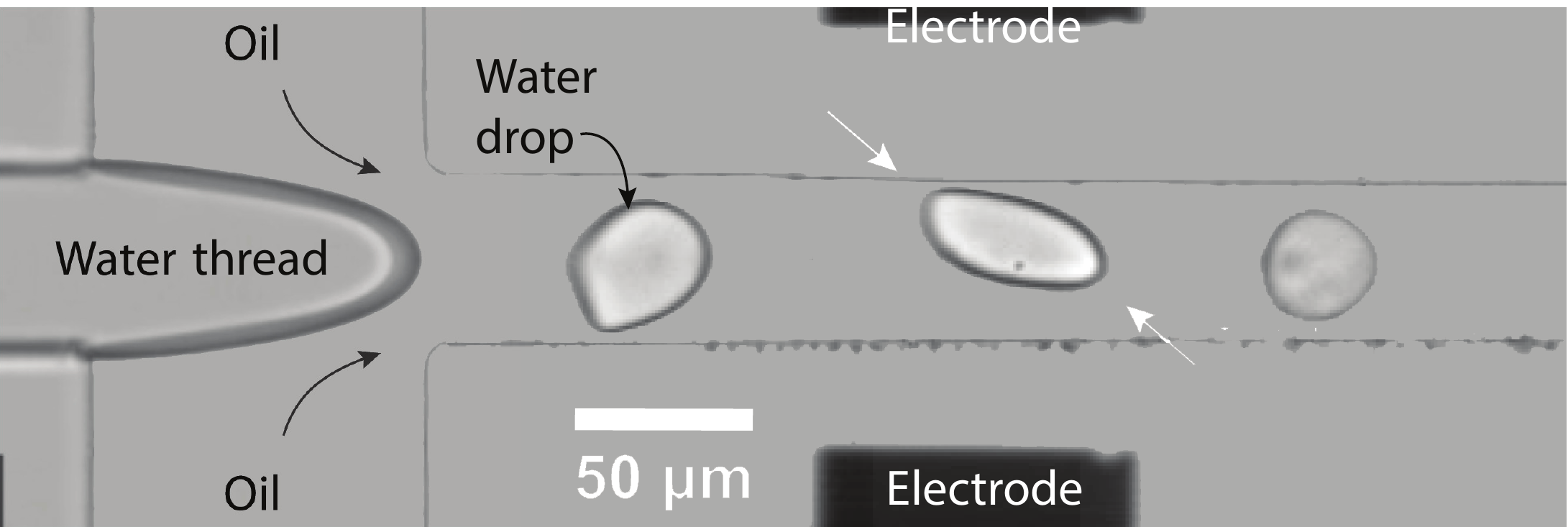}
\end{center}
\caption{Water droplets in the microchannel as they pass by the inductor electrode (black at bottom and top). Their shape becomes elliptical with singular tips (marked with white arrows) that tend to emit daughter droplets, identically as in the classical Rayleigh explosions [Videos added as supplementary material].\label{fig2}}

\end{figure}

\begin{figure}
\begin{center}
\includegraphics[width=0.4\textwidth]{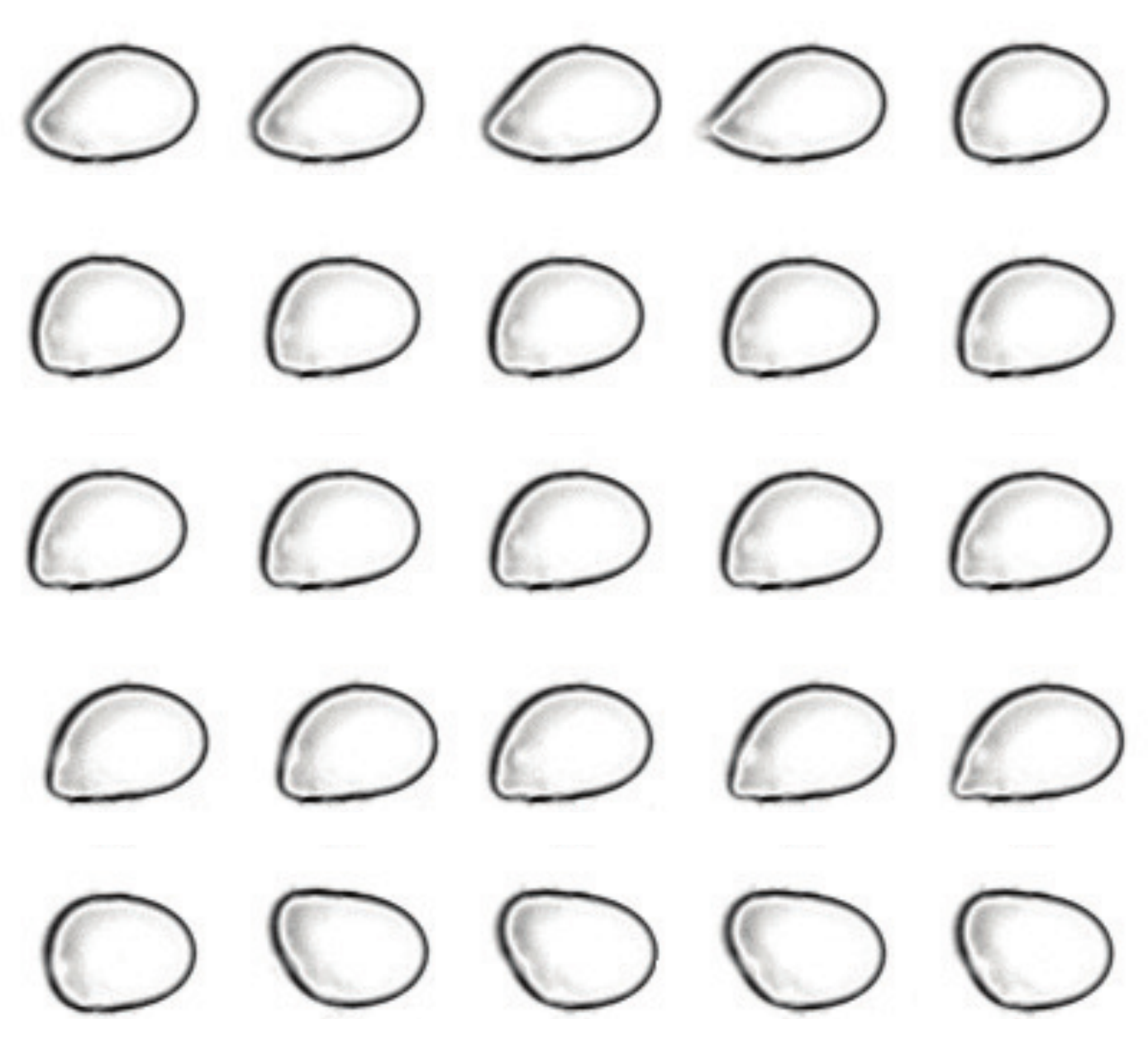}
\end{center}
\caption{Sequence of images depicting the deformation and relaxation of a droplet as it passes to its closest position from the inductor electrode. The time between frames is approximately 28 microseconds. The outer-to-inner flow rate ratio is 8.7, yielding a drop frequency of about 1 kHz. The video is included as supplementary document. \label{fig3}}

\end{figure}

The liquids employed are silicone oil Rhodorsil 47 v20 (density $\rho=950~kg/m^3$, viscosity  $\mu=20~cSt$) as external phase, and a salty water solution as inner liquid ($0.9\%$ NaI, electrical conductivity circa $5~ mS/cm$). The droplet generation frequency is controlled by the oil-to-water flow rate ratio. At low frequencies, the droplets flow normally without any special feature. As the frequency is increased, the charge relaxation time becomes comparable to the typical break-up time, and the charge buildup is then triggered at a certain critical frequency. After a typical charging time of a few seconds, the droplets lose their apparent circular shape and develop transient tips and microscopic jets as they pass close to the inductor electrodes, as shown in figures \ref{fig2} and \ref{fig3}. In spite of the high rate of droplet generation, the Reynolds number, defined as $Re=\rho U D/ \mu$, where $U$ is the typical external flow velocity and $D$ the typical droplet size, is kept at values below $10^{-2}$. Deformation and break-up of droplets by electric fields is a classical topic which gave birth to what we know nowadays as  Electrohydrodynamics. The pioneering works by Lord Rayleigh \cite{Rayleigh:1882} and later by Taylor and Melcher \cite{taylor1964disintegration, Melcher:1969vh} contain all the elements needed to understand the origin of these complex phenomena. The topic has continued developing in the present \cite{Leisner:2003,FernandezDeLaMora:2007fl,Collins:2008eq}, mainly due to its applications in mass spectrometry \cite{Fenn1989}. When an isolated droplet with an electrical charge $Q$ is subjected to a constant electric field $E$, the charges migrate to the poles of the droplet (dictated by the direction of the field). The charges exert a normal stress that tends to deform the droplet elliptically. When the field reaches a critical value, singular tips appear at the tip of the drops, which are regularized by charge and mass emission on the form of smaller and highly charged daughter droplets. Such an event occurs at critical values of the electrical field $E_c$ which is in general a function of the surface tension of the liquid, its electrical permittivity and its size. The maximum charge bearable for an isolated charged droplet in the absence of electrical fields ($Q\neq 0, E=0$) was solved analytically by Rayleigh \cite{Rayleigh:1882}. Recently, Fontelos et al. \cite{Fontelos:2008ct} solved numerically the transient problem for viscous and conducting drops, being able to calculate the critical electrical field as a function of the droplet charge, and identifying all the different deformation modes.

% \section { Calculations for the system }

In the case of the Kelvin water dropper, it is interesting to note that the voltage in the inductor electrode is unavoidably linked with the charge that the droplet acquires in an early stage before pinching-off. We make use of this to find a relation between the charge carried by a droplet and the electric field present in our microfluidic system. COMSOL has been used to solve the electrostatic problem considering that the jet is grounded, the inductor electrode is at 1\,V and the droplet charge is equal to the charge induced on the jet surface (see supplementary material for details). From the solution of the electric potential, the charge on the droplet and the maximum magnitude of the electric field on its surface ($E_{\mathrm{max}}$) can be obtained. Following the formalism by Fontelos et al. \cite{Fontelos:2008ct}, we define dimensionless charge and electrical fields respectively as $X=Q^2/(32\pi^2\varepsilon_{oil}\gamma R^3)$ and $E^*=E\sqrt{\varepsilon_{oil} R/\gamma}$, where $R$ is the droplet radius, $\varepsilon_{oil}$ is the electrical permittivity of oil and $\gamma$ is the oil-electrolyte interfacial tension. The following relation is found for our system:

\begin{equation}\label{relation}
E^*_{\mathrm{max}}=2.52\sqrt{X}
\end{equation}

\begin{figure}
\begin{center}
\includegraphics[width=0.5\textwidth]{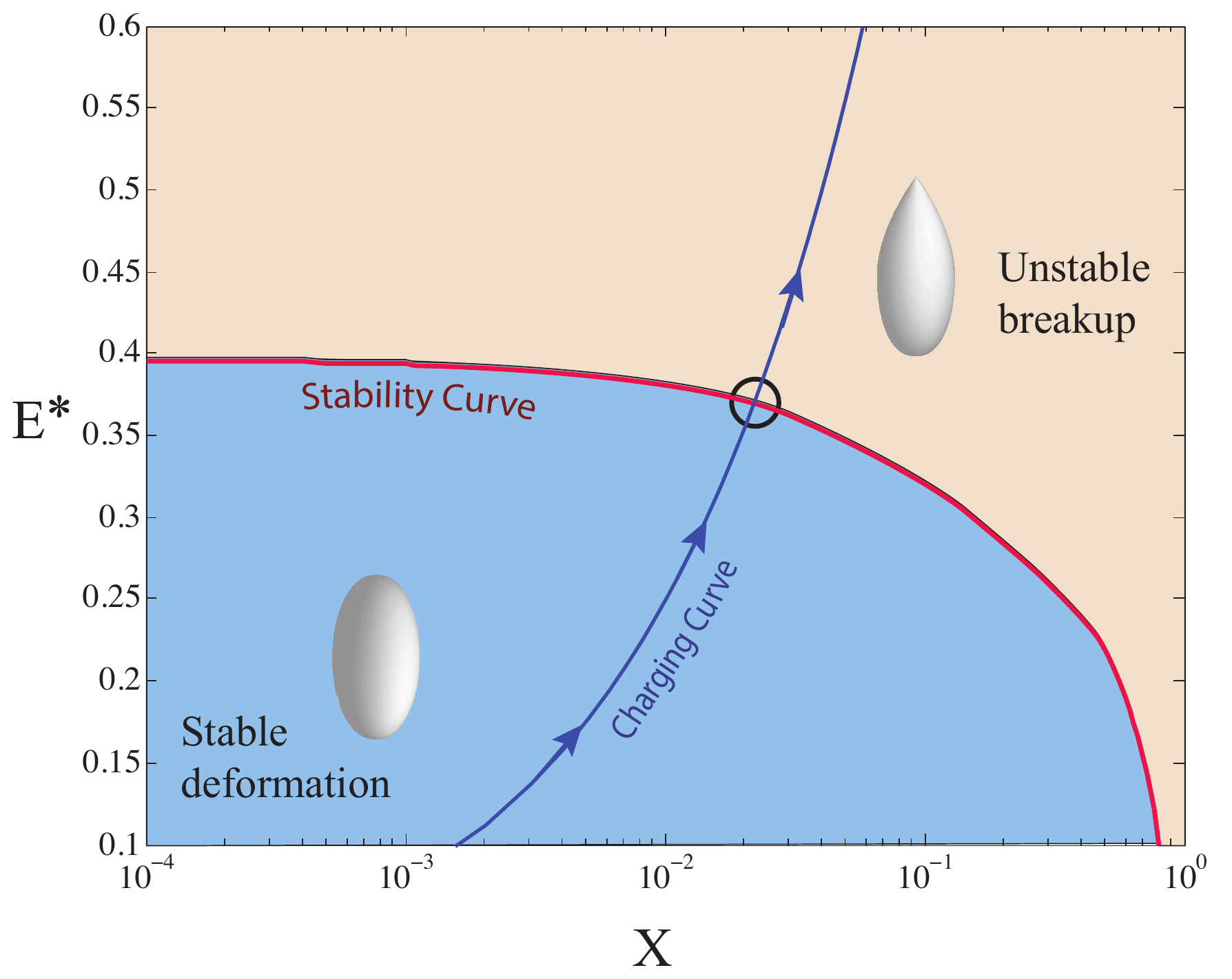}
\end{center}
\caption{Droplet Stability diagram in terms of the dimensionless electric field $E^*$ between droplet and inductor electrodes and the dimensionless droplet charge $X$. The curve in red represents the stability droplet limit calculated by Fontelos et al. \cite{Fontelos:2008ct}. The blue curve represents the charging curve followed by the geometry employed in our experiments $E^*= 2.52\sqrt{X}$. The droplets get charged following the blue curve until droplet instability occurs at $E^*=E_c^*$. Eventually, the electrodes will get discharged and the process starts again.\label{fig4}}

\end{figure}

Where the factor $2.52$ is specific of our geometry, but the scaling $E^*_{\mathrm{max}}\sim\sqrt{X}$ is universal. In figure \ref{fig4} we plot the stability curve calculated via numerical simulations by Fontelos et al. \cite{Fontelos:2008ct} for the case of an homogeneous electrical field, and the calculated $E^*_{\mathrm{max}}(X)$ curve (eq. \ref{relation}) obtained with our electrostatic simulations. We acknowledge that the situation in our system is different of that in Fontelos et al., where an homogeneous electric field is applied. But for the sake of comparison between both situations, we take $E_{\mathrm{max}}$ as a representative value of the electric field magnitude around the droplet.

In a normal working procedure, the system will get quickly charged, and the electric field experienced by the droplets will increase following the $E^*(X)$ curve until $E^*$ it reaches the critical stability value $E_c^*$ (red curve in figure \ref{fig4}). When this stage is reached, the droplets emit their charge in the form of tiny droplets before reaching the collector electrode and the system gets discharged again. This critical point actually gives us the maximum droplet charge achievable before the instability is triggered, which in our particular design and droplet size yields values of several kilovolts for the inductor electrode voltage and 0.2 pC of droplet charge. Such values have been experimentally checked by introducing an additional electrode downstream the inductor electrode connected to an electrometer. When a charged droplet passes by, it induces a charge in the electrode proportional to its own charge that can be measured with the electrometer. The average charge in time was then measured with a precision electrometer (Keithley electrometer 6517B) and with low-current ammeters (Keithley picoammeter 6485), yielding maximum electrical currents in the order of 0.5 nA.

%Following this approach we were able to measure the rising \chapter{¥}
%rge until maximum values of $2\pm0.5~pC$.

% \section{ Future Improvements }

The presented analysis gives a good understanding of the phenomena taking place in the system, but it would indeed require further improvements in the future. It would be necessary to recalculate the stability curve from Fontelos et al. \cite{Fontelos:2008ct} assuming a non-homogeneous electric field as that depicted in figure S3 in the supplementary material. This is however beyond the scope of this paper; we only note that the non-homogeneities of the electrical field may modify the presented curve substantially depending on the electrode configuration. It would also be necessary to take into account sharp corners, roughness, etc, which could induce locally extremely large electrical fields. Regarding the hydrodynamics, viscous stresses in the external flow around the droplets could also play an important role even at low Reynolds numbers, as noted by Fontelos et al. \cite{Fontelos:2008ct} and could explain the oblique angles at which the droplets burst respect to the electrode position.

% \section{ Conclusions & Future technology }

%
%\begin{figure}
%\begin{center}
%\includegraphics[width=0.4\textwidth]{figs/fig4new.pdf}
%\end{center}
%\caption{a) Computational domain showing the water thread, the inductor electrodes and a droplet. Droplet radius is 45\,$\mu m$ and the other dimensions
%are taken from the experimental device. b) solution of the electric potential in the microfluidic system (top view).}
%\label{fig4}
%\end{figure}

\balance
Depending on the desired application, different electrode and fluidic channel configurations can be designed to promote higher charging of the droplets without break-up (higher $X$ with $E^*<E_c^*$), or droplet break-up at low charging (low $X$ with $E>E^*$, the current case). The phenomenon is not only a interesting way of charging droplets but also to provoke their break-up. In many applications, the breakup or the electroporation of cells or vesicles is required to allow access to them  \cite{dimova2007giant}. For example, the current device could be able to provoke continuous breakup of targeted capsules. The present geometry contains only the basic elements for this phenomenon to occur, but it could be easily combined with other features, like different branches, or with several inductor electrodes. Finally, it would be interesting to explore the phenomenon to transform hydraulic pressure into electricity in an efficient way. Multi-component water droppers as the one proposed by Markus Zahn \cite{Zahn:1973wj} could be employed to multiply the energy outcome, which is now easily achievable with the severe miniaturization of the microfluidic devices.

\footnotesize{
%\bibliography{thunderstorm} %your .bib file
\providecommand*{\mcitethebibliography}{\thebibliography}
\csname @ifundefined\endcsname{endmcitethebibliography}
{\let\endmcitethebibliography\endthebibliography}{}

\bibliographystyle{rsc} }

\end{document}